x
\documentclass[12pt]{iopart}
\usepackage{amssymb}

\newcommand{\lb}{\label}
\newcommand{\var}{\varepsilon}
\newcommand{\vE}{{\bf{E}}}
\newcommand{\om}{\omega}
\newcommand{\al}{\alpha}

\newcommand{\pa}{\partial}

\newcommand{\fr}{\frac}
\newcommand{\de}{\delta}

\newcommand{\bw}{\begin{widetext}}
\newcommand{\ew}{\end{widetext}}
\newcommand{\be}{\begin{equation}}
\newcommand{\ee}{\end{equation}}
\newcommand{\bee}{\begin{equation*}}
\newcommand{\eee}{\end{equation*}}
\newcommand{\ba}{\begin{eqnarray}}
\newcommand{\ea}{\end{eqnarray}}
\newcommand{\bal}{\begin{align}}
\newcommand{\eal}{\end{align}}
\newcommand{\bml}{\begin{multline}}
\newcommand{\eml}{\end{multline}}
\newcommand{\non}{\nonumber}

\newcommand{\vk}{{\bf k}}

\newcommand{\vp}{{\bf{p}}}
\newcommand{\vB}{{\bf{B}}}

\newcommand{\vev}{\bf{v}}

\newcommand{\Nv}{{\mathbf{N}}}

\def\cF{{\cal F}}

\def\vk{{\bf k}}
\def\vr{{\bf r}}
\def\cF{{\cal F}}

\def\e{{\rm e}}
\begin{document}
\title[Plasma graviton production in TeV-scale gravity] {Plasma graviton production in TeV-scale gravity}

\author{ E Yu Melkumova }

\address{Department of Physics, Moscow State University, 119899,
Moscow, Russia} \ead{elenamelk@srd.sinp.msu.ru}

\begin{abstract}
We develop the theory of interaction of classical plasma with
Kaluza-Klein (KK) gravitons in the ADD model of TeV-scale gravity.
Plasma is described within the kinetic approach as the system of
charged particles and  Maxwell field both confined on the brane.
Interaction with multidimensional gravity living in the bulk with
$n$ compact extra dimensions is introduced within the linearized
theory. The KK gravitons emission rates are computed taking into
account  plasma collective effects through  the two-point
correlation functions of the fluctuations of the plasma
energy-momentum tensor.  Apart from known mechanisms (such as
bremsstrahlung and gravi-Primakoff effect) we find essentially
collective channels such as the coalescence of plasma waves into
gravitons which may be manifest in turbulent plasmas. Our results
indicate that commonly used rates of the KK gravitons production  in
stars and supernovae may be underestimated.

\end{abstract}

\section{Introduction}
TeV-scale gravity proposal due to Arkani-Hamed, Dimopoulos and Dvali
(ADD) \cite{ADD99} suggests that the standard model particles live
in the four-dimensional subspace (the brane)  of the D-dimensional
bulk with $ n = D - 4 $ extra dimensions compactified on a torus
which are inhabited only by gravity. The D-dimensional Planck mass
$M_*$ is supposed to be TeV-scale, and the large extra dimensions
(LED) to have sub-millimeter size. In this scenario there is an
infinite tower of massive KK gravitons \cite{HLZ99} whose existence
may be detected using table-top and collider experiments,
astrophysical and cosmological observations. It has been pointed out
that one of the strongest bounds on the parameters come from the
supernova SN1987A data \cite{ADD99}-\cite{SS08}. Processes
contributing to energy loss due to graviton emission from SN1987A,
the red giants and the Sun include photon-photon annihilation,
electron-positron annihilation,  nucleon-nucleon gravitational
bremsstrahlung, gravi-Compton-Primakoff scattering
\cite{GP99}-\cite{SS08}.\par
  Previous calculations of the KK graviton emission rates were
performed using one-particle Feynman rules with subsequent averaging
over Bose-Einstein and Maxwell-Boltzmann distributions to get the
results reliable for finite temperature $T$ \cite{BHKZ99}. These
calculations, however, did not take into account plasma collective
effects such as Debye screening, interaction of charged particles
with plasma waves etc, which may be important in astrophysical
conditions. This is the goal of the present contribution. We
generalize kinetic theory of interaction of plasmas with
gravitational waves developed earlier in 3 + 1 dimensions
\cite{GGM83} to the ADD model and present calculation of
gravi-bremsstrahlung rate from non-relativistic thermal isotropic
electron-ion plasma.

\section{Kinetic approach}
Consider collisionless plasma consisting of charged particles of
types $\al$ with parameters $e_\al,\,m_\al$ described by
microscopic distribution function \cite{Sit67,Sit73}
 \be
\cF_\al(x,\vp)=\sum_{i=1}^{N_\al}\de(\vr-\vr_i(t))\de(\vp - \vp_i
(t)),\label{mdf}
  \ee
 normalized as
 $  \int\cF_\al(x,\vp)d{\bf p} \,\,d^3 x=N_\al, $ satisfying the
 kinetic equation
 \be
 \fr{\pa \cF_\al}{\pa t}+\vev
\fr{\pa \cF_\al}{\pa \vr} + e_\al (\bf{E}+[\vev\bf{B}])\fr{\pa
\cF_\al}{\pa \vp}=0.  \lb{cf}
 \ee
They interact via the electromagnetic field $\bf{E},\,\bf{B}$
satisfying Maxwell equations
    \be  {\rm div} \vE=4\pi\rho, \quad
 {\rm rot}\vE=-\fr{\pa \vB}{\pa t},\quad
{\rm div}\vB=0,\quad {\rm rot}\vB=4\pi\bf{j} +\fr{\pa \vE}{\pa t},
\lb{max}\ee with the source terms
 \be \label{sorce}\rho(x)=\sum_\al
e_\al\int\cF_\al(x,{\vp})d{\bf p},\,\;\;\;{\bf{j}}(x)=\sum_\al
e_\al\int\fr{{\bf p}}{p_\al^0}\cF_\al(x,{\vp})d{\vp}.
 \ee To describe
gravitational radiation we will need the 3-spatial components of the
plasma energy momentum tensor $$ T_{ij}={}_mT_{ij}+{}_F T_{ij},$$
where
 \ba &&{}_mT_{ij}=\sum_\al \int \fr{p_i
p_j}{p_\al^0}\cF_\al(x,\vp)\, d{\bf p},\quad p_\al^0=\sqrt{{\bf
p}^2+m_\al^2},\quad x\equiv\{t,\bf{r}\},\label{tp}\\
 &&{}_F T_{ij}=-\fr{1}{4\pi} \left(
E_iE_j+B_iB_j-\de_{ij}\fr{E^2+B^2}{2}\right).\label{tf}\ea

To solve the system of equations (\ref{cf}-\ref{sorce})  we use
perturbation theory in terms of the electric charges. First we
separate fluctuations from the average distribution using the
approach of
 \cite{GGM83,Sit67}:
\be \cF(t,{\vr},{\vp})=f^0_\al+\de f_{\al}^0+\de f_\al,\lb{f}\ee
 where
$f^0\equiv <\cF_\al(x,{\vp})>$ - the  equilibrium Maxwell
distribution function: \be f_\al^0=\fr{N_{0\al}}{(2\pi
v_{T_\al}^2)^{\fr32}}e^{-\fr12\fr{v^2}{v_{T_\al}^2}},\quad
v_{T_\al}=\sqrt{\fr{T_\al}{m_\al}} , \quad v_{T_\al}<<1, \lb{max}\ee
$\de f^0_\al$- represents fluctuations due to chaotic particle
motion ("zero" fluctuations), while $\de f_\al(t,{\vr},{\vp})$ -
stands for fluctuations due to electromagnetic interaction of the
particles. Using the fact that $f_\al^0$ satisfies the free equation
(for zero charges), we obtain: \be \fr{\pa \de f_\al}{\pa t}+{\vev
}\fr{\pa \de f_\al}{\pa \vr} +e_\al {{(\bf{E}+[\vev\bf{B}])}}\fr{\pa
( f^0_\al+\de f_{\al}^0+\de f_\al )}{\pa \vp}=0. \lb{cfg}\ee We then
further expand the fluctuation $ \de f_\al$ in power series in terms
of charges: $ \de f_\al =\de f_\al^1 + \de f_\al^2 + ... $, and use
the Fourier-transformation $$ f(x)=\fr{1}{(2\pi)^4}\int
f(k)\e^{-ik_\mu x^\mu}d^4k $$ to get in the first two orders
  \ba\de f^1_\al({{k},{\vp}})=-\fr{i
e_\al}{(\om-\vk\vev)}{\bf{F}}(k)\fr{\pa f^0_\al}{\pa \vp} ,\qquad
{\bf F}\equiv\bf{E}+[\vev\bf{B}],\lb{f1}\ea
 \ba \fl\de f^2_\al({k,\vp})=-\fr{i
e_\al}{(2\pi)^4(\om-\vk\vev)}\int{\bf F}(k-k_1)\fr{\pa}{\pa
\vp}(\de f_\al^0 (k_1,\vp)+\de f_\al^1 (k_1,\vp))d^4k_1.\lb{f2}\ea
 The corresponding expansion of the charged particles  energy-momentum
 tensor  will read: \ba
{}_mT_{ik}={}_mT^0_{ik}+\de {}_mT^0_{ik}+\sum_{l=1}^\infty  \sum_\al
\int \fr{p_i p_j}{p_\al^0}\de f_\al^l(x,\vp)\, d{\bf p}.\lb{t2}\ea
First non-zero contribution to gravitational radiation comes from
terms of the second order in the fields $\bf{E},\,\bf{B}$, namely $
\de T_{ik}=\de\, {}_mT_{ik}^2+\de \,{}_FT_{ik}$.
\section{Gravitational radiation}
In  ADD model one considers \cite{HLZ99} the linearized $D=4+n$
dimensional Einstein equations
  $$
\Box{\psi}{\!\!}_{MN}=\varkappa^2_D T{\!\!}_{MN}, $$ where
$\psi_{MN}=h_{MN}-\eta_{MN}h_P^P/2,\;\; h_{MN}=g_{MN}-\eta_{MN}$,
the indices $M,\,N$ run over the brane $\mu,\,\nu=0,1,2,3$ and $n$
directions on the torus, and the harmonic gauge
$\partial_N\psi^{MN}=0$ is understood. Formula for the total
(integrated over space and time) energy loss on gravitational
radiation in ADD model  was recently derived in \cite{GKST10}:
 \ba
 \fl
{\cal{E}}=\frac{\varkappa_D^2}{16 \pi^{3} V_d}\sum_{\Nv \in
\mathbb{Z}^d} \int d\vk T_{SN}(k) T^*_{LR}(k)\tilde{\Lambda}^{SNLR}
\left. \vphantom{\sqrt{d}}
    \right|_{k^0=\sqrt{|{\vk} |^2+(2 \pi \Nv /L)^2}},
    \quad V_d=(2\pi L)^{D-4},
\ea where $T_{SN}(k)$ is the four-dimensional Fourier-transform, and
\ba\label{gr perts11aADD}
\tilde{\Lambda}^{SNLR}=\frac{1}{2}\left[\eta^{SL} \eta^{
NR}+\eta^{SR} \eta^{ NL}\right]-\frac{1}{D-2}\eta^{SN} \eta^{ LR}
\ea is the polarization projection operator.
 Since the stress tensor has only brane indices $\mu=0,i$ and it
satisfies the conservation equation $k_\mu T^{\mu\nu}(k)=0$, one can
eliminate its components $T^{0\nu}$ in favor of $T^{ij}$. The
resulting expression will contain the sum of various order terms
from which the lowest non-zero contribution  comes from the product
of the second order terms. We also have to normalize the energy loss
of the stationary plasma per unit volume and per unit time, which is
achieved by introducing the two-point correlation function according
to the relation
$$ < T_{\al\beta}(k) T^*_{\al'\beta'}(k+\Delta k)>  =  <
T_{\al\beta} T^*_{\al'\beta'}>_k \delta^4(\Delta k).$$ The final
expression for the gravitational emission rate to all graviton modes
per unit plasma volume per unit time will read\ba &&\fl
P=\frac{\varkappa_D^2}{16 \pi^{3} V_d}\sum_{\Nv \in \mathbb{Z}^d}
\int d{\vk} <\de T_{ik} \de T^*_{i'k'}>_k\tilde{\Lambda}^{iki'k'}
\left. \vphantom{\sqrt{d}}
    \right|_{k^0=\sqrt{|{\vk} |^2+(2 \pi \Nv /L)^2}},
\lb{lastADD2}\\ &&\fl
\tilde{\Lambda}^{iki'k'}=\frac{1}{2}\left[\Delta^{ii'} \Delta^{
kk'}+\Delta^{ik'} \Delta^{ i'k}\right]-\frac{1}{D-2}\Delta^{ik}
\Delta^{ i'k'} , \quad \Delta_{ik}=\de_{ik}-n_in_k, \quad
{\bf{n}}=\fr{{\vk}}{{k^0}}.\lb{LA}\ea In our approach we did not use
explicit decomposition of the full multidimensional metric
perturbation $h_{MN}$ in massless and massive modes, these
correspond to $\Nv=0$ and $\Nv\neq 0$ terms in the sum respectively.
For massless modes the three-dimensional vector ${\bf{n}}$ is the
unit vector, but not for massive modes. Note that $h_{MN}$ has bulk
components due to the trace term in the wave equation.

To calculate gravitational bremsstrahlung from the non-relativistic
plasma we need the longitudinal field and the related fluctuation
function. Keeping as the source in Maxwell equations terms of the
zero and the first order
 \ba {\rm div} {\vE}=4\pi e\int d{\vev} (\de f^0+\de f^1),\quad
  \de f^1=-i \fr{e}{m}\fr{1}{\om-\vk\vev}{\vE_{k}}\fr{\pa f^0}{\pa {\vev}},\lb{divE}\ea
 one can express the  first order electric field through  zero
fluctuations:
   \ba &&\vE_{k}=-4\pi
 i\fr{\vk}{|{\vk}|^2}\fr{\rho^0_{\vk\om}}{\var_L(k )},\qquad
  {\rho^0_{k}}=\sum e\int d{\vev }\de f^0_{k}(\vev)\lb{Erho}, \\
 && \var_L(k)=1+\fr{4\pi e^2}{m\om |{\vk}|^2}\int d{\bf{p}}
    {\fr{(\vk \vev)}{(\om - \vk\vev )}}{\vk}\fr{\pa f_0(\vev )}{\pa \vev
    }\lb{epL},\ea the last line being  the longitudinal permittivity of the isotropic
plasma.
 Using this, one then finds the two-point correlation functions
defined above \cite{Sit73}.\footnote{
 Density fluctuations can be characterized by correlation functions
whose Fourier-transforms exhibits homogeneity and isotropy:
 $<\de n(\vr_1,t_1)\de n(\vr_2,t_2)>\equiv <\de n^2>_{\vr,t},\quad
 \vr=\vr_1-\vr_2,t=t_1-t_2.$
 The space-time Fourier transformation will give the spectral fluctuation density:
  $<\de n^2>_{k}=\int d\vr\int d t \e^{-i\vk\vr+i\om\vr}<\de
  n^2>_{\vr,t},$
 $<\de n(k)\de n(k')> =(2\pi)^4<\de n^2>_{k}\de^4(k - k').$}
Correlation function for zero-order particle density  $<\de
n^2>^0_{k }$  is calculable for the Maxwell distribution (\ref{max})
in the finite form, while the field correlation functions can be
then expressed through this quantity. Using the equations
(\ref{divE}) and (\ref{Erho}) one find the following relations
between $<\rho^2>_{k}$,
  $<E^2>_{k}$ and ${<\rho^2>^0_{k}}$:
  \ba \fl<\rho^2>_{k}=\fr{<\rho^2>^0_{k}}{|\var_L(k)|^2},\,\,
<E^2>_{k}=\fr{16\pi^2}{|{\vk}|^2}\fr{<\rho^2>^0_{k}}{|\var_L(k)|^2},
\,\, <\de \rho^2>^0_{k}=\sqrt{2\pi}\fr{n_0
e_\al}{|{\vk}|v_T}\e^{-{\fr{\om^2}{2|{\vk}|^2v_T^2}}}. \lb{c99}\ea
Substituting this to (fluctuations of) the energy-momentum tensors
and making use of the formula   (\ref{lastADD2}), one obtains after
some rearrangements:
 \ba \fl<\de T_{ik}
\de T^*_{i'k'}>_k\tilde{\Lambda}^{iki'k'}=\non\\
\fl\int
 d^4k_1\left\{
A(k_1,k) \fr{T}{m}\fr{<\de n^2>^0_{k_2}}{|\var_L(k_2)|^2}<\de
n^2>^0_{k_1} + B(k_1,k) { \fr{<\de
n^2>^0_{k_1}}{|\var_L(k_1)|^2}\fr{<\de
n^2>^0_{k_2}}{|\var_L(k_2)|^2}}\right\}.
 \lb{TTL}\ea
  According to the relations (\ref{c99}), the first term here is
the product  of the field and particle density correlation
functions, while the second is quadratic in the field correlators.
The following frequency regions has to be distinguished: \par i)
Zero real part of the complex permittivity $ Re\var_L(\om,\vk)=0 $
in the denominators corresponds to propagation of the longitudinal
Langmuir waves with the frequency $$ \om\simeq\om_L ,\quad
\om_L=\sqrt{\fr{4\pi^2e^2N_0}{m}}
$$ and the long wavelength: {$ |{\vk}|<\fr{1}{r_D}$}, where {$
r_{D_e}=\fr{v_{T_e}}{\om_L}$} is the Debye shielding radius. \par
ii)
 For {$\omega > \om_L ,\quad |{\vk}|v_{T_e}\gg 1  $, then $ \var_L\simeq 1$}
 which corresponds to electromagnetic intaraction switched off.
 \par iii)
For {$ \omega < \om_L $ one  has $\var_L(k)\simeq
1+(|{\vk}|r_D)^{-2}\quad
$} which corresponds to the Debye shielding.\\
 Correspondingly, we have to distinguish radiation losses due to
gravitational bremsstrahlung (denominators are non-zero, both terms
being taken into account), and the coalescence of two Langmuir
plasmons into graviton ( the second term with both  denominators
 at Langmuir frequencies).
\subsection{Gravitational bremsstrahlung} This corresponds to  $Re
\var_L\neq 0$. The main contribution to the integral comes from the
region $\om < |{\vk}_1|
 v_{T_e}$, so for $ k_0 >\om_L $, $|{\vk}_1|r_D >1$, then $ \var_L
 \simeq 1 $. In opposite limit $ k_0<\om_L $, Debye shielding
becomes manifest, then $\var_L(k_1)\simeq 1+(|{\vk}_1|r_D)^{-2} $.
 Therefore we obtain in the two cases:
  \ba
P\simeq\frac{128}{\sqrt{\pi}}(1+\sqrt{2}) e^4 { N_0^2}{v_{T_e}}
\fr{T^{n+1}}{M_*^{n+2}}
 \Big\{I_1(n)\ln{\Lambda}+\fr{8}{3}I_2(n)
 \Big\},\quad k^0>\om_L,\lb{ppp1}\ea
 \ba P\simeq\frac{128}{\sqrt{\pi}}(1+\sqrt{2}) e^4 {
N_0^2}{v_{T_e}} \fr{\om_L^{n+1}}{M_*^{n+2}}
 I_1(n)\ln{\Lambda},
 \quad k^0<\om_L,\lb{ppp2}\ea
where $ \ln{\Lambda} $ is the Coulomb
 logarithm, $\Lambda={r_{D_e}}/{\hat{r}} $,
 $ r_{D_e}=\fr{v_{T_e}}{\om_L} $ -- is the
 electron Debye radius,   $\hat{r}$ is determined from the condition $
  {e^2}/{\hat{r}}=T $ and $M_*=\fr{1}{\varkappa_D^2}$ is the D-dimensional
   Planck mass. The coefficients
 $ I_1(n)$, $I_2(n)$ depend on the number of extra
 dimensions $n,\;2\leq n\leq 7$  as follows:
  ($
   I_1(n)=1.1,\,\,1.07\,\,0.97\,\,0.83,\,\,0.67, \;\;
   I_2(n)=4.8,\,\,6.35,\,\,7.3,\,\,7.5,\,\,7.1
   $.
 If $ n=0 $ ($ I_1(n)=1 $, $ I_2(n)=1 $) these results coincide with those - in $ 3+1 $ dimensions
 \cite{GGM83}.
Without collective effects, the result  in $ 3+1 $ dimensions
corresponds to the quadruple formula calculation   with consequent
thermal averaging
 \cite{W65}.
\subsection{Coalescence of two Langmuir plasmons into graviton}
For coalescence of two Langmuir plasmons into the graviton the field
correlation function can be presented as
 \ba <E^2>_{k} = 4\pi^2 T(\de(\om-\om_L)+\de(\om+\om_L))\lb{e2}\ea
 The graviton frequency will be twice the Langmuir
frequency and the rate of emission will be
  \be
 P^{0}=2e^4N_e^4v_{T_e}\fr{(2\om_L)^{n+1}}{M_*^{n+2}}I_n,\lb{co}\ee
with $ I_n=1.4,\,\,1.9,\,\,2.2,\,\,2.3,\,\,2.2, $ for
$n=2,\ldots,6$.

Compare with the result of the ref. \cite{BHKZ99} for the
bremsstrahlung losses which was obtained using the one-particle
approach:\ba P\simeq
\sum_j\fr{\Gamma(\fr52+n)n_en_jZ^2_j\al^2I_{GB}(n)}{\Gamma(\fr32)}
\fr{T^{n+1}}{M_*^{n+2}},\lb{Bar}\ea where $ n_e $ and $ n_j $ are
the electrons and ions densities, $Z_j$ is the ions charge,
$\al=e^2$, and  the numerical  integral $ I_{GB}(n) $ is presented
for $ n=2,3: \quad I_{GB}(n)=0.7,0.3 $.
 Being applied to  red giants  with the electron temperature $T \sim 8.6
$ keV  and the electron density $n_e= 3.0\cdot 10^{29} cm^{-3}$,
assuming the conservative upper limit on the energy-loss rates of
red giants $\quad \dot{\var} \sim 100 erg\, g^{-1} \,sec^{-1}\quad$,
the authors of \cite{BHKZ99} obtained the upper limit for Planck
mass $ M_* $ to be  $10^3$ for $n=2$ and $  1.7\cdot 10^{-5}$ for
$n=3$. Our result (\ref{ppp2}) for the bremsstrahlung losses is
about three time greater than (\ref{Bar}). Moreover, an additional
enhancement comes from the in
 coalescence process (\ref{co}).\par
 The work was supported by RFBR project 08-02-01398-a.


\end{document}